\documentclass[runningheads]{llncs}

\usepackage{orcidlink}
\usepackage[misc]{ifsym}

\usepackage{xcolor}
\usepackage{graphicx}
\usepackage{booktabs}
\usepackage{tabularx}
\usepackage[strings]{underscore}
\usepackage{subfig}

\usepackage{xcolor}
\hypersetup{
    colorlinks=true,
    linkcolor={blue!100!black},
    citecolor={blue!100!black},
}

\newcommand{\squeezeup}{\vspace{-2.5mm}}

\usepackage{hyperref}

\begin{document}
\title{Edge AI-Based Vein Detector for Efficient Venipuncture in the Antecubital Fossa\thanks{Accepted for publication in MICAI 2023, Part II, LNCS 14392 }}
\titlerunning{Edge AI-Based Vein Detector for Efficient Venipuncture}
\author{Edwin Salcedo\inst{1}\orcidlink{0000-0001-8970-8838} \and
Patricia Peñaloza\inst{1}\orcidlink{0009-0004-2151-485X}}
\authorrunning{E. Salcedo and P. Peñaloza}

\institute{Department of Mechatronics Engineering\\
Universidad Católica Boliviana “San Pablo”, La Paz, Bolivia 
\email{\{esalcedo,patricia.penaloza\}@ucb.edu.bo}}

\maketitle              
\begin{abstract}

Assessing the condition and visibility of veins is a crucial step before obtaining intravenous access in the antecubital fossa, which is a common procedure to draw blood or administer intravenous therapies (IV therapies). Even though medical practitioners are highly skilled at intravenous cannulation, they usually struggle to perform the procedure in patients with low visible veins due to fluid retention, age, overweight, dark skin tone, or diabetes. Recently, several investigations proposed combining Near Infrared (NIR) imaging and deep learning (DL) techniques for forearm vein segmentation. Although they have demonstrated compelling results, their use has been rather limited owing to the portability and precision requirements to perform venipuncture. In this paper, we aim to contribute to bridging this gap using three strategies. First, we introduce a new NIR-based forearm vein segmentation dataset of 2,016 labelled images collected from 1,008 subjects with low visible veins. Second, we propose a modified U-Net architecture that locates veins specifically in the antecubital fossa region of the examined patient. Finally, a compressed version of the proposed architecture was deployed inside a bespoke, portable vein finder device after testing four common embedded microcomputers and four common quantization modalities. Experimental results showed that the model compressed with Dynamic Range Quantization and deployed on a Raspberry Pi 4B card produced the best execution time and precision balance, with 5.14 FPS and 0.957 of latency and Intersection over Union (IoU), respectively. These results show promising performance inside a resource-restricted low-cost device. The full implementation and data are available at: \url{https://github.com/EdwinTSalcedo/CUBITAL}

\keywords{Vein detection \and Deep learning \and NIR Imaging \and Edge AI}
\end{abstract}
\section{Introduction}

\begin{figure}[ht]
  \centering
    \centering
  \subfloat[NIR light penetration through skin layers until reaching the subcutaneous tissue where veins and arteries locate.]{\includegraphics[width=0.48\textwidth]{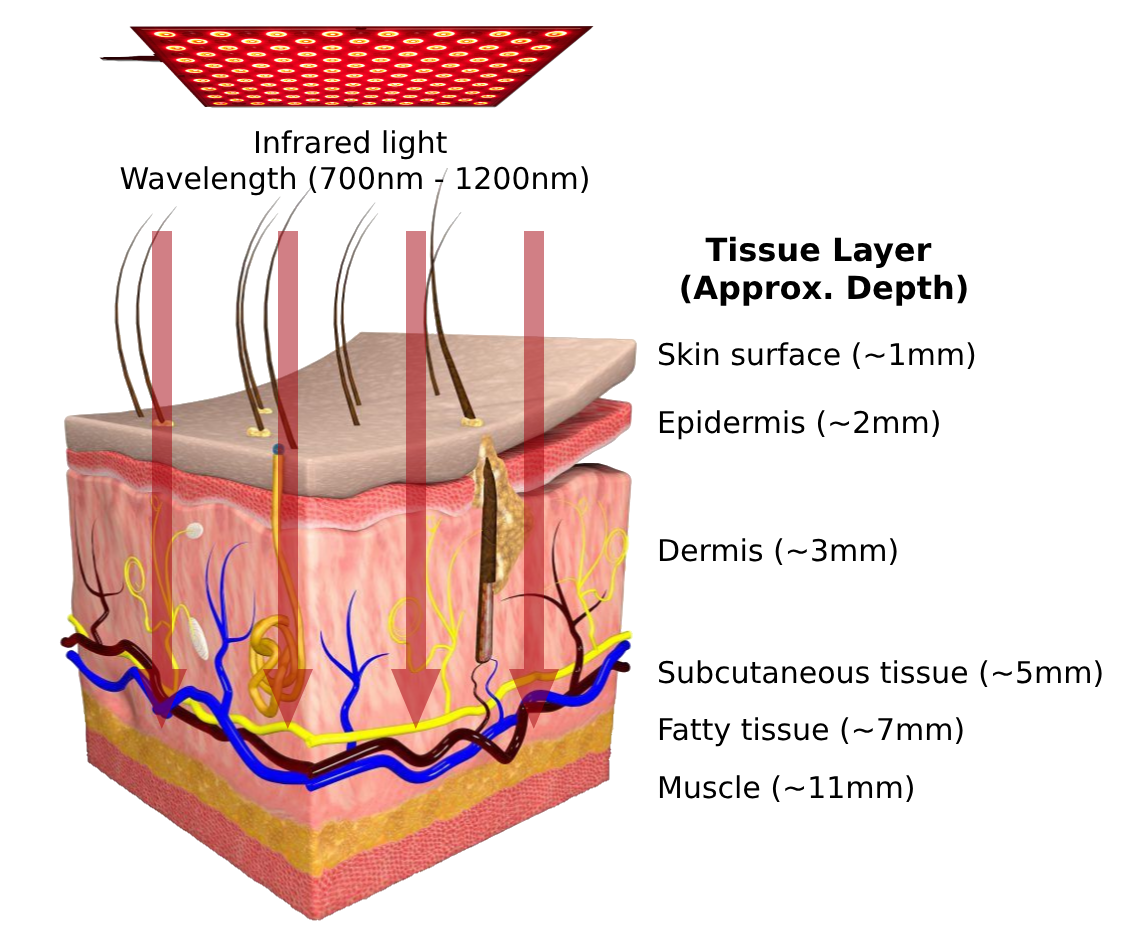}\label{fig:tissue-layer}}
  \hfill
  \subfloat[Samples of vein distributions in the arm region with the antecubital fossa marked in green (adapted from \cite{corzo2014}).]{\includegraphics[width=0.48\textwidth]{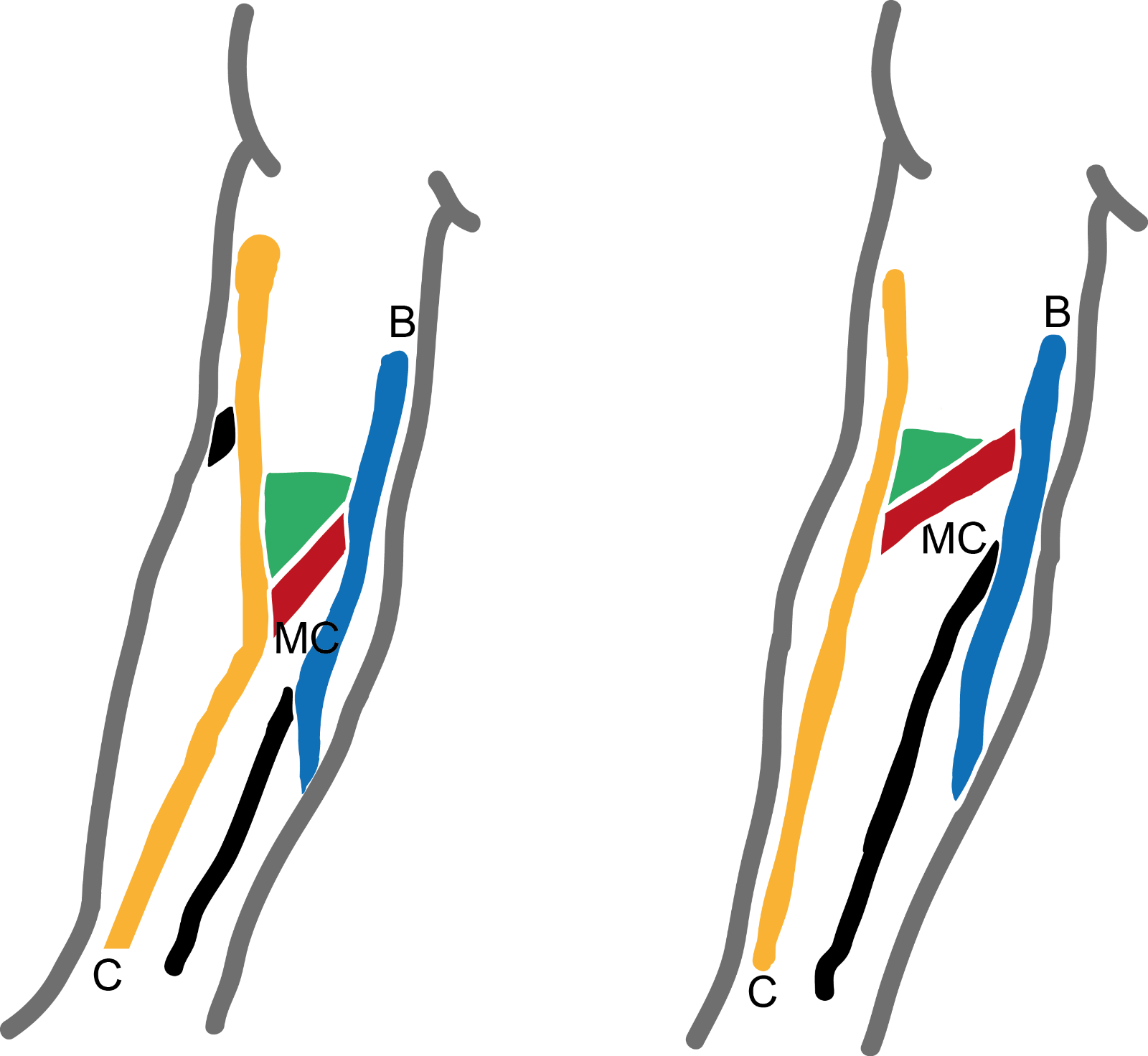}\label{fig:vein-distribution}}
  \caption{Anatomy of forearm veins in the antecubital region.}
\end{figure}

Venipuncture is a necessary procedure applied by medical staff, either to draw a blood sample, start an intravenous infusion, or instil a medication. While this procedure can be applied to several regions of the anatomy, doctors prefer the antecubital fossa due to the higher visibility and stability of veins there. Initially, physicians identify and ascertain suitability of the median cubital (MC), cephalic (C) and basilic (B) veins in the antecubital fossa, as depicted in Figure \ref{fig:vein-distribution}. It is worth mentioning that the median cubital vein is usually referred as the best site to perform catheterization \cite{naik2019}\cite{corzo2014}. However, people who do not have good vein visibility might require longer pre-inspection times, which can cause an early start of a trial-and-error venipuncture process to localize a suitable vein. This is the case for children, elderly people, dark-skinned people, and people with overweight or diabetes. Palpation, warm water, tourniquets, NIR vein finders are among some well-known good practices to improve vein visibility. Yet, if veins are still not noticeable, the need for health professionals to assist the next patients might cause bruises, pain, and bleeding to the current one.

\squeezeup
\squeezeup

Since the beginning of the 2010s, several companies started commercializing hand-held vein finders based on ultrasound, transillumination, or infrared light to facilitate venipuncture. Nowadays, these devices' features range from basic vein visualization enhancement to simultaneous detection and mapping of veins in any part of the body (e.g. AccuVein AV400 and AV500). However, the widespread adoption of these devices has been rather limited owing to their high cost and closed software. Recently, in response to these limitations, several proposal systems based on Computer Vision, Deep Learning, and Near Infrared imaging (NIR) have emerged as promising approaches for vein visualization enhancement \cite{chaoying2021}\cite{zaineb2022}\cite{francisco2021}\cite{chen2021}. Nevertheless, they are usually designed to improve vein visualization in the entire forearm region, so healthcare professionals must still choose the most suitable region or vein with which they should work. Also, most recent algorithms are oriented to run in a central server, instead of being deployed to portable devices. So, there is still room for research to develop better AI-based devices that recommend which vein or region to select for venipuncture in real-time and on-site.

Deep learning at the edge can be applied not only for more precise NIR imaging-based vein segmentation, but also to identify which region to choose for venipuncture. Therefore, our proposal aims to extend this body of work with the following contributions:

\begin{itemize}
  \item A new dataset containing 2,016 NIR images with low visible veins in arms is introduced, in tandem with their respective ground truth vein segmentation masks. The dataset also comprises bounding box, centroid and angle annotations for antecubital fossa localization inside the images.
  \item We test five DL-based semantic segmentation models and perform a thorough comparison, from which we select and modify the best one to also act as a regression model for antecubital fossa localization and arm direction prediction. 
  \item We test the resulting model on four common microcomputers (Raspberry Pi 4B, Raspberry Pi 3B+, Khadas VIM3, and NVIDIA Jetson Nano) and using four common quantization modalities (dynamic range quantization, full-integer quantization, integer quantization with float fallback, and float16 quantization). The best combination is finally implemented in a bespoke, portable device that shows suitable veins in the antecubital fossa.  

\end{itemize} 

The remainder of the paper is structured as follows. Section \ref{sec:literature-review} presents the state of the art on vein image acquisition approaches, as well as new DL and Edge AI-related tendencies for vein localization. Section \ref{sec:material-and-methods} describes the prototyping process of the end device as well as the implemented DL models and metrics. Then, in Section \ref{sec:experimental-results}, we present the experimental results in terms of prediction accuracy and inference time. Finally, Section \ref{sec:conclusion} offers conclusions and discusses potential future research threads.

\section{Literature Review}
\label{sec:literature-review}

Many image acquisition, processing, and visualization techniques have been proposed and released to the market to enhance subcutaneous vein localization. By way of illustration, AccuVein vein finders feature simultaneous localization and mapping using light projections towards the skin. Nevertheless, their prices range from 1,800 USD to 7,000 USD per unit \cite{aimvein2022}, which keeps them inaccessible to many medical centers in developing countries. In the current section, we present a review on the main technologies and research trends on open-source vein detectors development. 

\subsection{Image acquisition approaches}

Two image acquisition approaches can be clearly distinguished for forearm vein localization: transillumination-based and reflectance-based methods. The first ones are more extended in the literature because of their portability and low-cost. They mainly transmit light through the skin and tissue of a body sector, which is then followed and captured by a light sensitive camera at a given wavelength. While regular RGB cameras capture light in the human visible spectrum (400-700 nm), transillumination-based techniques such as multi-spectral imaging or hyper-spectral imaging aim to capture illumination in different ranges of the electromagnetic spectrum, e.g. the ultraviolet range or the infrared range. This approach was widely explored by investigators. For instance, Shahzad et al. \cite{Shahzad2013} propose an illumination wavelength selection algorithm for vein detection using a multi-spectral camera, such that the system can recommend what wavelength to use for a patient based on his skin-tone.

\squeezeup
\newcolumntype{Y}{>{\centering\arraybackslash}X}
\begin{table*}[htbp]
    \caption{Summary of recent forearm vein distribution detector proposals from 2018 to present.}\label{tab:state-of-the-art}
    \begin{center}
        \begin{tabularx}{\textwidth}{YYYYY}
        \toprule
        \textbf{Year \& Ref.} & \textbf{Imaging method} \& \textbf{Camera} & \textbf{Detection method} & \textbf{End device} & \textbf{Key metrics} \\ 
        \midrule
        2022 \cite{Leipheimer2022} & NIR \& US & U-Net & PC & 0.83 IoU \\
        2022 \cite{jongwon2022} & NIR, Pi NoIR 2 & Image Processing & PCB \& VideoCore-IV & 74.93 \% SSIM\\
        2022 \cite{kuthiala2022} & NIR, OV5647 Omnivision & U-Net & Raspberry Pi 4B & 0.68 DSC  \\
        2022 \cite{zaineb2022} & NIR, JAI \& DALSA X64-CL & Pix2Pix & & 0.96 DSC \\
        2021 \cite{chaoying2021} & NIR \& RGB, JAI \& DSLR & FCNN & Nvidia Jetson TX2 & 0.78 Accuracy \\
        2021 \cite{chen2021} & NIR \& US, Pi NoIR 2 \& US Probe & semi-ResNext-U-Net & Raspberry Pi 4B & 0.81 DSC \\
        2018 \cite{surya2018} & NIR \& Pi NoIR & Image Processing & Raspberry Pi 2 & 0.84 Accuracy \\
        \bottomrule
        \end{tabularx}
    \end{center}
\end{table*}
\squeezeup

Particularly, Near-Infrared light (NIR) has been broadly explored over the past years as a vein visualization enhancing technique. As shown in Figure \ref{fig:tissue-layer}, this requires NIR illumination and a special camera able to capture NIR transillumination, which in turn generates digital images. NIR light can go through human skin reaching between 700 nm and 1,200 nm depth depending on the person's complexion. Since this range can provide information on a body's temperature and structure, it makes it suitable to capture vein presence in the subcutaneous tissue. Furthermore, oxygenated and deoxygenated hemoglobin, two components of blood, absorb and transmit NIR light better through them. About NIR capture devices, some common cameras available in the market are described in Table \ref{tab:state-of-the-art} (under the ``Imaging method \& Camera'' column) from where we can conclude Raspberry Pi NoIR 1 and 2 are the most frequented NIR cameras for research. For instance, academics in \cite{Danni2016} proposed a detection device that combines two NIR cameras to obtain depth information about the subcutaneous layer of an arm and overlapped 3D visualizations of veins to enhance their illustration.

Ultrasound imaging (US) and photoacoustic imaging are amongst the most-used methods in reflectance-based commercial devices for forearm vascular localization. While US provides a high-resolution frame-of-reference for identifying density, flow and perfusion of veins, Photoacoustic Imaging (PI) permits registering important factors such as oxygen saturation, total hemoglobin and the microdistribution of biomarkers. Both solve the problem of finding vessels by reflecting a high frequency sound (US) or non-ionizing light (PI) over a focused part of a body. Then, the return time travel of the reflected waves is registered with an imaging probe as electrical signals \cite{Fronheiser2010}. These waves, also known as ultrasonic waves, are detected by ultrasonic transducers to reconstruct physiological organs in living beings. In the case of human vessels, hemoglobin concentration and oxygen saturation are physiological properties that form 2D or 3D images with distinguishable contrast between skin tissue and vessels due to their distance concerning the light source. Combining both US and NIR modalities has recently brought new opportunities for robotic catheterization. For instance, researchers in \cite{chen2021} employed both NIR and US imaging inside a robot to perform venipuncture autonomously. A similar combination of US and NIR imaging was brought to a handheld robotic device by Leipheimer et al. in \cite{Leipheimer2022}, where the authors propose the use of machine learning models to safely and efficiently introduce a catheter sheath into a peripheral blood vessel.

\subsection{Computer-based vein distribution localization}

Venipuncture success of intravenous procedures depends on the timely localization of veins. Although a great majority are applied in the antecubital fossa, some procedures require finding veins in lower arm sections. Thus, semantic segmentation of veins over the forearm region is a crucial task that should be performed as precisely and timely as possible. Specifically, semantic segmentation aims to classify each pixel inside a collected image with a label. Most investigations interpret veins anatomically as hollow tube structures that join each other along an image, and they assume two categories for each pixel: vein pixel and background pixel. There are two notorious computer vision-based approaches that are regularly applied for forearm vein segmentation: traditional image processing methods and deep learning architectures. 

\subsubsection{Image processing-based methods}

Segmentation approaches based on traditional image processing methods for NIR, US, multi-spectral, and hyper-spectral images usually comprehend steps for contrast and illumination enhancement, morphological operations, vein structure discovery, and edge detection. For instance, several investigations apply Histogram Equalisation or Contrast-Limit Adaptive Histogram Equalisation to enhance the contrast of the input images \cite{yildiz2019} \cite{azueto2017}. Then, vessel segmentation approaches aim to discriminate regions with veins from the background. Here, vein segmentation techniques can be also classified as vein structure-based, region-based, gradient-based, and pixel-based. For example, Li et al. \cite{li2017} proposed a convex-regional-based gradient preserving method that use edge information to enhance the low contrast and reduce the noise in NIR images for better vein segmentation. By applying a convex function, they find global minimums as optimal locations to detect veins. Recently, researchers in \cite{jongwon2022} proposed an image preprocessing system for existing vein detection devices to remove hair digitally from NIR images. They achieved an improvement of 5.04\% of Structural Similarity Index (SSIM) with respect to their original vein segmentation algorithm, which shows the relevance of image processing methods for newer approaches.  

\subsubsection{Deep learning-based methods}
Recently, deep learning has demonstrated huge success in detection tasks from visual information due to its generalization power. Recent investigations leverage deep learning-based algorithms to classify pixels as vein or background inside the collected images. In contrast to image processing, deep learning models do not require strict controlled environments, which makes them more suitable to perspective, distance or illumination variations. U-Net based architectures are amongst the most used approaches for vein semantic segmentation \cite{kuthiala2022} \cite{chen2021}. Moreover, Shah et al. \cite{zaineb2022} proposed a forearm vein segmentation model based on the Pix2Pix architecture to translate NIR images of arms into their segmented vascular versions. Their architecture consists of a student, which is a U-Net model that learns to generate new vascular masks from NIR images, and a teacher, which is a PatchGAN-based model that discriminates each generated image into fake or original images. Combined into a common architecture, the approach obtained 0.97 of accuracy. This model outperformed previous methods for forearm vein segmentation.

\subsection{Edge AI methods}
During the last years, the advent of better microprocessors has increased the opportunities to bring deep learning models into standalone end devices. Edge AI and Edge Computing are two paradigms that have attracted much of the attention recently. While Edge Computing aims to bring information processing closer to the users, Edge AI is the implementation of artificial intelligence in an edge computing environment. Edge AI-based environments are usually implemented in embedded systems beside Computer Processing Units, Graphics Processing Units (GPUs), Tensor Processing Units (TPUs), Vision Processing Units (VPUs), Field Programmable Gate Arrays (FPGAs), Application-Specific Integrated Circuits (ASICs), or Systems-on-Chip (SoC) \cite{chaoying2021}. Specifically, development cards such as Nvidia Jetson, Khadas, Neural Computer Stick, or Google Coral can be of huge help to speed up new Edge-AI based applications. 

Hardware implementation and Edge AI are important for the present project since venipuncture is applied to patients in situ. So far, several investigations have proposed forearm segmentation algorithms deployed in portable scanners or fixed stations, and the host devices range from Raspberry Pi cards \cite{kuthiala2022} \cite{chen2021} \cite{kuthiala2022} \cite{surya2018} \cite{yildiz2019} to NVIDIA Jetson cards \cite{chaoying2021}. An updated publication list on vein distribution finders is described in Table \ref{tab:state-of-the-art}. To the best knowledge of the authors, the great majority of investigations focus on general forearm vein segmentation and do not detect specific sites for optimal venipuncture. For instance, Chaoying et al. \cite{chaoying2021} proposed one of the first investigations to deploy deep learning-based forearm vein detectors on an embedded device with meaningful results: 0.78 of accuracy and 0.31 seconds per processed frame. 

The availability increase of low-cost NIR cameras, such as Pi NoIR, Jai, OV5647, Omnivision, among others, make them suitable for new on-device forearm vein finding applications. For example, Ng. et al \cite{ng2023} proposed a vein detection and visual guidance system to show the location of veins through a mixed-reality-based interface. They used a HoloLens 2 device and its infrared emitter to obtain new images, which in turn let them segment and visualize veins in real time. Their vein segmentation approach was based on the U-Net architecture with a RegNet-based encoder and achieved 0.89 precision. In the case of robotized venipuncture, Chen et al. \cite{chen2021} proposed a robotic system solution named Venibot to determine an optimal area on a forearm and perform puncturing autonomously. Their proposal combines US and NIR imaging to control the movements of the venipuncture robot. 

\section{Material and Methods}
\label{sec:material-and-methods}

To localize the hidden veins of a patient, we developed a Deep Learning-based model that processes NIR images of their forearm and segments the present veins. This model also localizes the antecubital fossa to hide all veins except the ones located in that zone, such that a healthcare practitioner can only see the suitable veins for venipuncture. Later, the algorithm was implemented on an embedded system by applying compression techniques. In the present section, we describe the complete software and hardware implementation process in detail.

\subsection{Forearm Vein Segmentation}

\subsubsection{Dataset Collection and Preprocessing}

\begin{figure}[ht]
  \centering
  \includegraphics[width=0.7\linewidth]{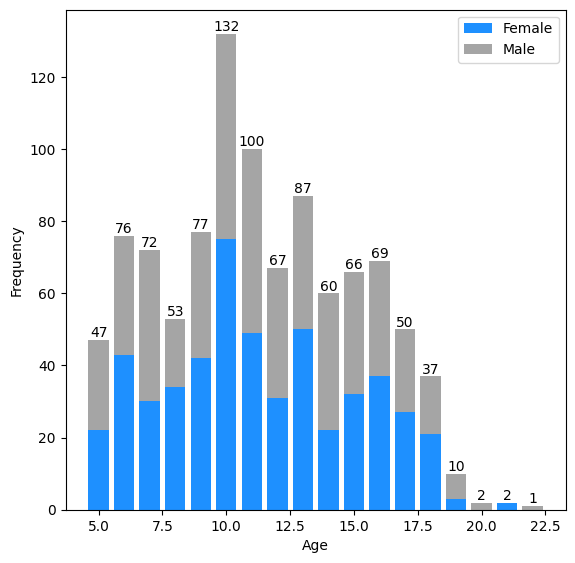}
  \caption{Age distribution of 1,008 subjects who volunteered in the data collection stage. }
  \label{fig:age-distribution}
\end{figure}

As stated before, dehydration, young and old age, overweight, dark skin tone, and diabetes are among some factors that can affect patients' veins visibility. Specifically, in the case of young subjects, this series of injections can cause medical trauma, which in turn might cause future self-medication and conflicting feelings when requiring healthcare assistance \cite{naik2019}. Therefore, the present research focused on enhancing the visibility of veins in young patients. 2,016 NIR images were collected from both arms of 1,008 young subjects during the year 2022. The volunteers, whose age frequency is shown in Figure \ref{fig:age-distribution}, were students in elementary and secondary schools in the cities of Sacaba and Santa Cruz, Bolivia. About the setup, each patient located an arm at a time on a flat surface covered with a white fabric for the sake of better contrast. Meanwhile, the initial version of the vein finder depicted in Figure \ref{fig:first-end-device} was located 30 centimeters above using a lamp arm printed in 3D. 

\squeezeup

Given that collecting information from children also requires parental consent, volunteers' parents were asked to sign a consent agreement to use the captured images for research purposes. This resulted in an approximate time of 5 minutes per subject, making a total of 83.8 hours. Data was saved and administered in a laptop's internal memory as CSV files and PNG images using a bespoke Tkinter application. This had the purpose of registering and managing NIR images along with the full name, complexion, age, medical condition, gender, and signed consent agreement of volunteers. To form the final version of the base dataset, NIR images were converted to grayscale and enhanced using Contrast-Limit Adaptive Histogram Equalisation (CLAHE). Then, ground-truth was manually annotated with background, arm, and vein segments using Roboflow. Finally, the images were normalized to 512 x 512 pixels to obtain pairs of images and masks suitable for semantic segmentation DL architectures.  

To avoid the risk of overfitting, we generated an augmented version of the base dataset applying sequential randomly-selected augmentation techniques. We implemented the following techniques from the ImgAug library: flipping images horizontally, perspective, rotating images in the range of $180^\circ$ and $-180^\circ$, blurring images with Gaussian and average filters, contrasting with gamma functions, among others. In the end, the augmented version of the dataset contained 8,000 images with their corresponding segmentation masks. 

\subsubsection{Model Selection and Training}

The recent progress made on vein subcutaneous segmentation based on NIR imaging in \cite{chaoying2021}\cite{zaineb2022}\cite{chen2021} let us understand the great generalization capabilities of Deep Learning-based (DL) methods with respect to previous approaches. Thus, we focused on implementing various recently proposed generic architectures for semantic segmentation:  U-Net, Segnet, PSPNet, DeepLabV3+, and Pix2Pix. The models were implemented using TensorFlow 2.12.0 and Colab Pro+ with NVIDIA A100 GPUs. Besides modelling with both tools, they let us code a unified data loading and munging pipeline for the dataset and experiment parallelly with multiple instances per model, so that optimizing the base code and hyperparameters was completed efficiently. Both versions of the dataset, the base one and augmented one, were split into three subsets: 70\% for training, 20\% for validation, and 10\% for testing. 

The available resources provided by Colab limited us to use a batch size of 8 instances per step when training each model. Although all models might have trained longer or shorter times, we made sure to use 10 epochs for a fair comparison. This was also supported by the fact that some models (DeepLabV3+ and Pix2Pix) started overfitting when training longer. We used Binary Cross Entropy (BCE) as the unique loss function for all models to measure the dissimilarity between the ground truth and predicted masks. A mathematical representation of BCE is shown in Equation \ref{eq:bce}, where \textit{$y_i$} and \textit{$\hat{y_i}$} represents a ground truth binary classification vector and a predicted binary classification vector, respectively. Also, in the same equation, {\em T } stands for the number of pixels per instance, and {\em f} for the sigmoid activation function, as defined in Equation \ref{eq:sigmoid}.

\begin{equation} 
\label{eq:bce}
BCE = -\frac{1}{T} \sum_{i=0}^{T} y_{i}\cdot log(f(\hat{y_{i}})) + (1 - y_{i})\cdot log(1 - f(\hat{y_{i}}))
\end{equation}

\begin{equation} 
\label{eq:sigmoid}
f(s_{i}) = \frac{1}{1+e^{-s_{i}}}
\end{equation}

Our Pix2Pix implementation was inspired on the work proposed by Zaineb et al. \cite{zaineb2022}, however, we followed the original Pix2Pix architecture proposed by Isola et al. \cite{Isola2017}. This base version
contained a generator based on the PatchGAN model and discriminator module based on the U-Net architecture. Therefore, we reused our base U-Net architecture and implemented PatchGAN. About the loss functions, we also applied BCE (as defined in Equation \ref{eq:bce}) to differentiate the ground truth and generated masks. Yet the generator required to use Mean Squared Error (MSE), commonly defined as in Equation \ref{eq:mse}, where $M$ is the number of image pairs (ground truth and predicted masks), and $N$ is the number of pixels per image pair. 

\begin{equation} 
\label{eq:mse}
\mathrm{MSE} = \mathrm{\frac{\sum_{i=1}^{N}(x_i-y_i)^2}{M*N}}
\end{equation}

To measure the models' precision, metrics such as Pixel Accuracy, Intersection over Union (IoU), Dice Score, Pixel F1Score, and Peak signal-to-noise ratio (PSNR), were calculated according to Equations \ref{eq:pixel-accuracy}, \ref{eq:iou}, \ref{eq:dice}, \ref{eq:f1score}, and \ref{eq:psnr}, respectively. These metrics required to process ground truth $G$ and predicted $P$ masks first, with which we quantified the well and wrong classified pixels as True Positive ($TP$), True Negative ($TN$), False Positive ($FP$), or False Negative ($FN$), as defined in \cite{wrojas2023}. 

\begin{equation}
\label{eq:pixel-accuracy}
  \mathrm{Accuracy} = \frac{\mathrm{TP} + \mathrm{TN}}{\mathrm{TP} {+}\mathrm{FP} {+}\mathrm{TN}{+}\mathrm{FN}}
\end{equation}

{\small
\setlength\tabcolsep{0pt}
\noindent\begin{tabular*}{\textwidth}{@{\extracolsep{\fill}}
p{.50\textwidth}
p{.50\textwidth}
@{}}
  \begin{equation}
  \mathrm{IoU} = \frac{|G \cap P|}{|G \cup P|}
    \label{eq:iou}
  \end{equation} &
  \begin{equation}
  \mathrm{Dice Score} = \frac{\mathrm{2*|P \cap G|}}{\mathrm{|P|} {+} \mathrm{|G|}}
    \label{eq:dice}
  \end{equation}
\end{tabular*}
}

{\small
\setlength\tabcolsep{0pt}
\noindent\begin{tabular*}{\textwidth}{@{\extracolsep{\fill}}
p{.50\textwidth}
p{.50\textwidth}
@{}}
  \begin{equation}
  \mathrm{F1Score} = \mathrm{\frac{2*TP}{2*TP+FP+FN}}
    \label{eq:f1score}
  \end{equation} &
  \begin{equation}
  \mathrm{PSNR} = \mathrm{10*log_{10}(\frac{ 255^{2}}{MSE})}
    \label{eq:psnr}
  \end{equation}
\end{tabular*}
}

\begin{figure}[ht]
  \centering
  \includegraphics[width=\linewidth]{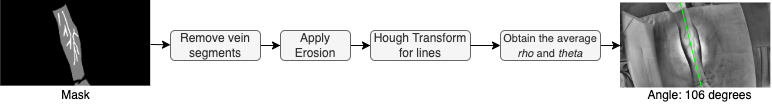}
  \caption{Angle extraction algorithm.}
  \label{fig:angle-extraction}
\end{figure}
\squeezeup

\vspace{-0.4cm}
\squeezeup

\subsection{Cubital Fossa Localization}

Once we found U-Net was the best semantic segmentation architecture for the task, we continued the investigation by experimenting with methods to localize the Antecubital Fossa. This required another labelling iteration to enclose the cubital fossa region with a bounding box in all 2,016 NIR images on Roboflow. Moreover, we made sure the bounding boxes' centroids were exactly located in the fossa, which means the center of the bounding boxes' coordinates were located in the median cubital (MC) areas in Figure \ref{fig:vein-distribution}. It is worth noting that the fossa location prediction also required the angle of the examined arm to hide veins out of the antecubital fossa region. So, we labelled the orientation of each arm synthetically by following the process shown in Figure \ref{fig:angle-extraction}. 

As depicted, we worked with the ground truth mask, removed veins, and converted the arm segment into a shape similar to a line by applying a series of morphological erosion operations. Then, we used OpenCV's function Hough Transform for Lines (HTL) to obtain the polar coordinates of lines from an accumulator matrix. According to this matrix, the more concur points in an image, the more probable they depict a line, therefore, HTL obtains a set of $\theta$ and $\rho$ where points frequently concur. Given that we started with a single line representing the entire arm, we averaged the $\theta$ and $\rho$ values and converted them to a degree value between 0 degrees and 180 degrees, starting from the very right in a counterclockwise direction. 

\begin{figure}[ht]
  \centering
  \includegraphics[width=\linewidth]{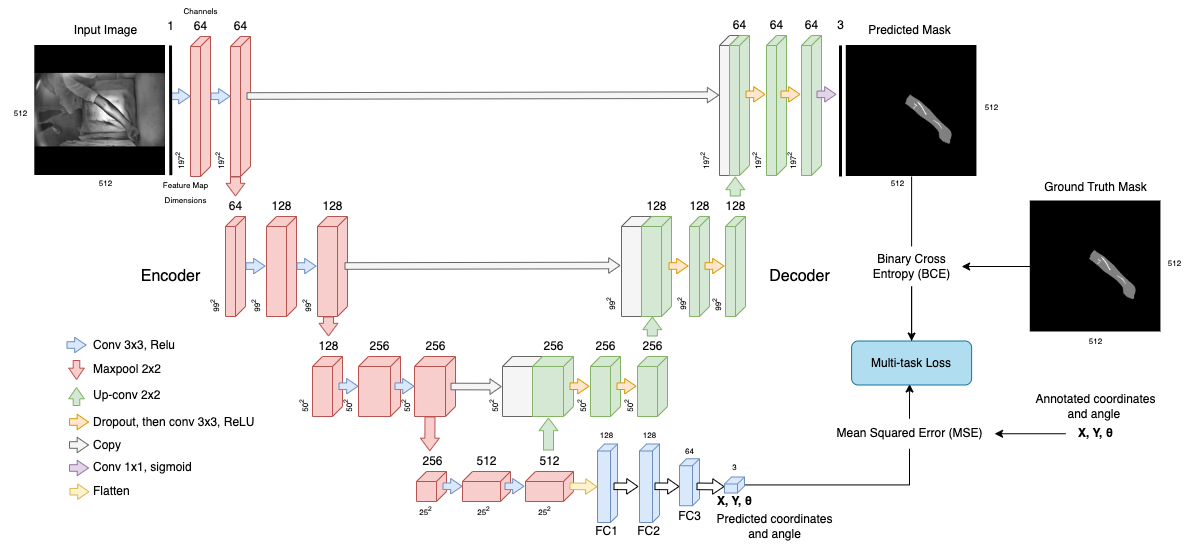}
  \caption{Final U-Net architecture implemented with TensorFlow for simultaneous forearm vein segmentation, forearm localization and arm angle detection.}
  \label{fig:final-unet}
\end{figure}

Obtaining the final version of the dataset let us model the problem as a combination of semantic segmentation and regression tasks. Thus, we integrated a neuronal network into the U-Net architecture. The layers, resolution, and channels of the final architecture are illustrated in Figure \ref{fig:final-unet}. Consequently, we created a multi-task loss function to combine BCE and MSE as defined in Equation \ref{eq:multitaskloss}. We also included the metric Mean Absolute Error (MAE) in the performance analysis stage when training and validating the architecture, as defined in Equation \ref{eq:mae}. 

\squeezeup

\begin{equation}
\label{eq:multitaskloss}
  \mathrm{MutiTaskLoss} = \mathrm{BCE} + \mathrm{MSE}
\end{equation}

\begin{equation}
\label{eq:mae}
  \mathrm{MAE} = \mathrm{\frac{\sum_{i=1}^{N}|x_i-y_i|}{M*N}}
\end{equation}

Finally, once the model was implemented and tested, we used the compression methods available in TensorFlow Lite to reduce the size of the model and embed it inside the final end device. The implemented approaches were Dynamic Range Quantization, Integer Quantization with Float Fallback, Full Integer Quantization and Float 16 Quantization.

\subsection{Hardware Development}

\subsubsection{Device design with its components}

\begin{figure}[ht]
  \centering
    \centering
  \subfloat[Initial imaging system]{\includegraphics[width=0.32\textwidth]{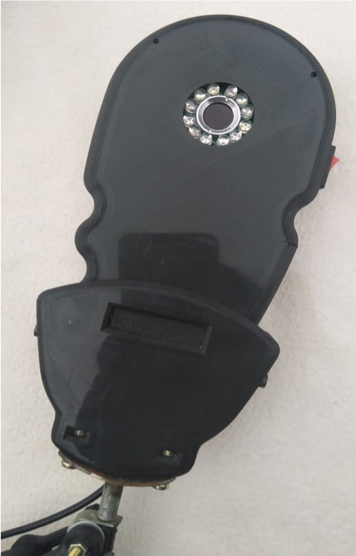}\label{fig:initial-end-device}}
  \hfill
  \subfloat[3D printed lamp-shape structure]{\includegraphics[width=0.57\textwidth]{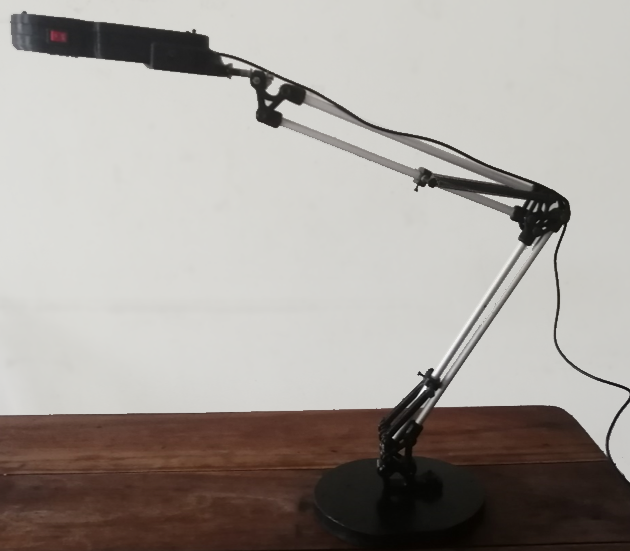}\label{fig:initial-end-device-lamp}}
  \caption{Initial prototype used to collect 2006 NIR images}
  \label{fig:first-end-device}
\end{figure}

The availability of 3D printing technology and standalone microcomputers has opened new possibilities for innovative product design and manufacturing. To prototype the vein finder, we integrated electronic circuitry design, components assembly and 3D printing techniques. Most importantly, the device required initially the implementation and parameter optimization of the NIR imaging system in order to improve the quality of acquired NIR images. The initial version of this system is shown in Figure \ref{fig:first-end-device}. 

\begin{figure}[ht]
  \centering
  \includegraphics[width=\linewidth]{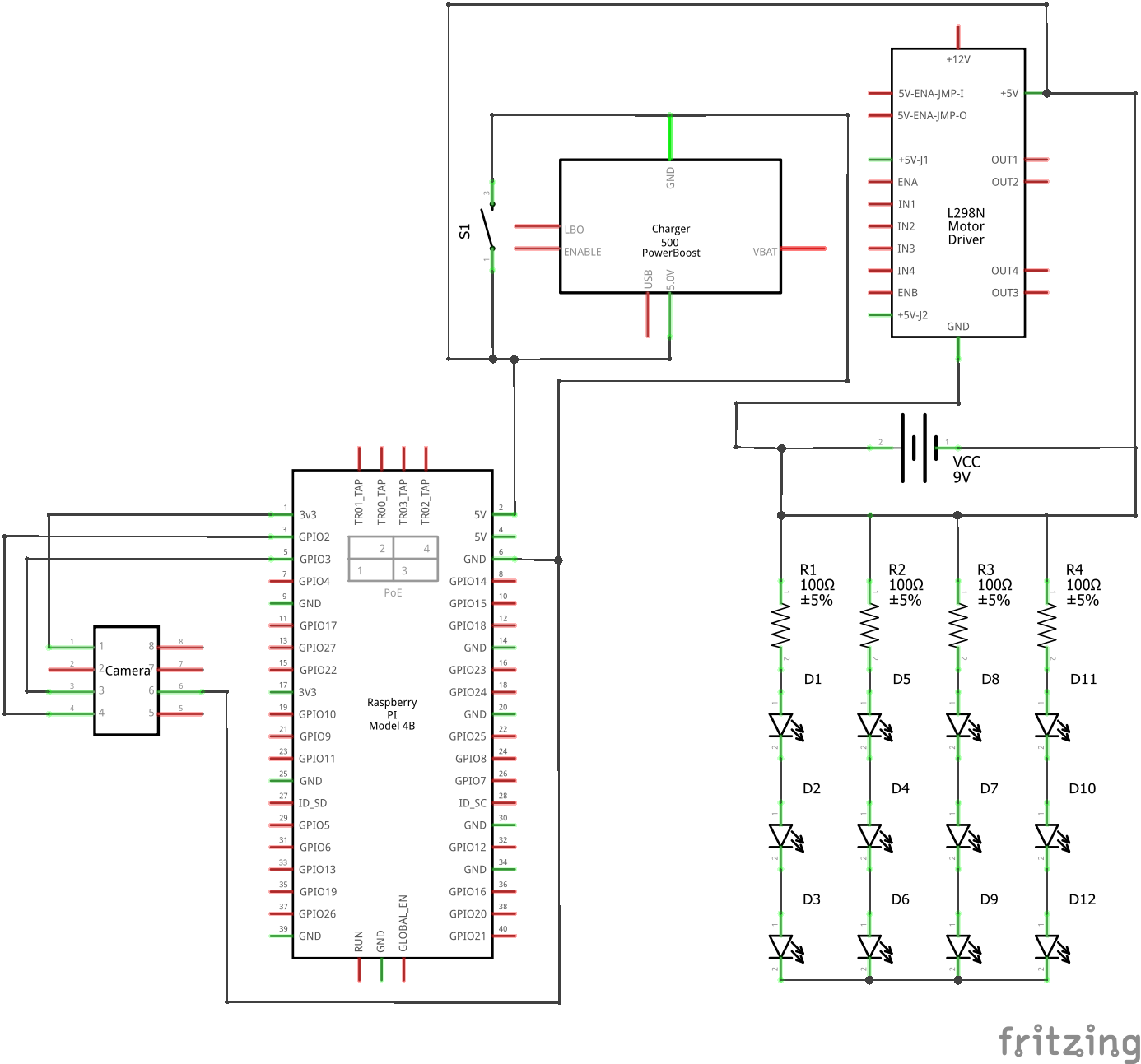}
  \caption{Electronics schematics}
  \label{fig:schematics}
\end{figure}

Moving on to the development of the final vein finder device, we aimed to develop an embedded system to contain a DL architecture for simultaneous vein segmentation and antecubital fossa localization. Then, the testing and compression stages of the final architecture carried out on different cards let us define that the Raspberry Pi 4B card was the best choice for the prototype due to its good balance with respect to cost, precision, and inference time. Consequently, it was chosen for on-device image processing and DL model deployment. To enhance portability and autonomy, a Xiaomi portable battery of 10000 mah was connected to the Raspberry Pi 4B card through a micro-USB cable. For image capture, we included a Raspberry NoIR V2 camera to the Raspberry Pi 4B through a 2-lane MIPI CSI camera port. A touch screen was also installed to provide a Graphic User Interface (GUI) to the end user. The electronics schematics are presented in Figure \ref{fig:schematics}.

The illumination matrix of 12 infrared LEDs developed in the initial prototype, as shown in Figure \ref{fig:first-end-device}, was included on a perforated breadboard, which was used to assemble the necessary circuits, mainly $100 \Omega$ $1/2$ W resistors. The LED matrix was powered by a $9$ V battery, considering an appropriate resistor for each group of 3 LEDs. In addition, a $5$ V relay module was implemented to control the energization of the LED matrix, and an On/Off switch was designed for the device activation. Moreover, a mechanism was implemented to synchronize the Raspberry Pi 4B and the relay module's power, ensuring simultaneous activation and deactivation.

\begin{figure}[ht]
  \centering
    \centering
  \subfloat[Isometric view]{\includegraphics[width=0.60\textwidth]{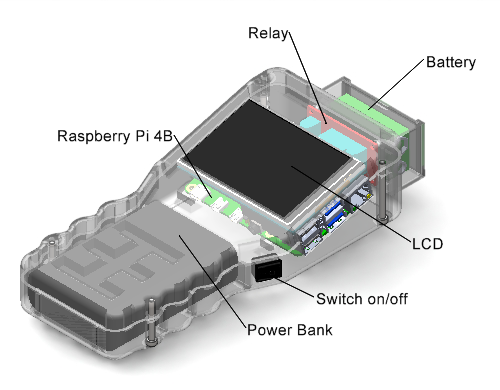}\label{fig:isometric}}
  \hfill
  \subfloat[Posterior view]{\includegraphics[width=0.35\textwidth]{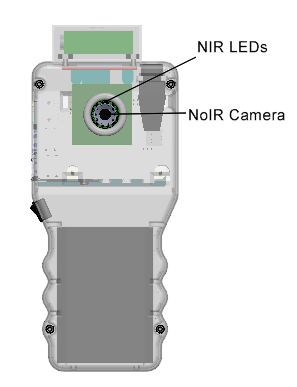}\label{fig:posterior}}
  \caption{Vein finder design using the 3D CAD software SolidWorks.}
  \label{fig:3Dcad}
\end{figure}

\subsubsection{Manufacturing and Assembly}

As presented in Figure \ref{fig:3Dcad}, the design and implementation of the external case aimed to embed the LED matrix, microcomputer card, battery, powerbank, and camera using 3D printing technology and Polylactic Acid filament (PLA). The top and bottom parts of the casing were printed separately, with careful control of printing parameters to achieve optimal layer adhesion and surface finish. In addition, the battery slot and camera cover were also printed as separate components to facilitate easy assembly and maintenance. The 3D-printed ergonomic case was designed with the software SolidWorks, which let us achieve a lightweight structure and sufficient rigidity, as well as durability and protection for the electronic components. Most importantly, the camera was mounted at the center of the illumination matrix, allowing for higher accuracy in vein detection under a frontal annular lighting setup. The final dimensions of the device are  23 cm x 9.5 cm x 3.5 cm. 

\section{Experimental Results}
\label{sec:experimental-results}

The present section reports the results of the two main detection tasks in the proposal: forearm vein segmentation, and antecubital fossa detection including forearm vein segmentation.

\subsection{Forearm Vein Segmentation}

A summary of the quantitative results of the models (Pix2Pix, U-Net, Segnet, PSPNEt, DeepLabV3+), calculating the metrics defined in Equations \ref{eq:pixel-accuracy}, \ref{eq:iou}, \ref{eq:dice}, \ref{eq:f1score}, and \ref{eq:psnr}, is shown in Table \ref{tab:segmentation-metrics}. The numbers in bold define the lowest or highest performance for each metric. While we aimed to obtain high values for almost all columns, we identified some models' weights (shown in the last column) were higher than others. This was an important factor in choosing one model over the others. For instance, we identified U-Net as one of the most precise models requiring fewer kilobytes than Segnet or Pix2Pix. 

\begin{table*}[htbp]
    \caption{Results comparison for forearm vein segmentation}
    \begin{center}
        \begin{tabularx}{\textwidth}{YYYYYYYYY}
            \toprule
             & \textbf{Model}  & \textbf{IoU} & \textbf{Dice Score} & \textbf{PSNR} & \textbf{Pixel Accuracy} & \textbf{F1-Score} & \textbf{FPS (Frames)} & \textbf{Weight (MBs)}\\ 
            \midrule
            \textbf{Base dataset} & U-Net & 0.986 &  0.050 & \textbf{70.050} & 0.992 & \textbf{0.992} & \textbf{5.940} & 1.600  \\
            & Segnet & \textbf{0.987} & 0.055& 70.010 & \textbf{0.993} & \textbf{0.992} & 4.290 & 2.100  \\
            & PSPNet & 0.948 & \textbf{0.516} & 63.590 & 0.969 & 0.967 & 5.630 & \textbf{1.100}  \\
            & DeepLab v3  & 0.981 & 0.120& 68.620 & 0.988 & 0.989 & 5.770 & 4.000 \\
            & Pix2Pix & 0.940 & \textbf{0.700} & 63.610 & 0.970 & 0.960 & 3.880 & 7.000 \\
            \midrule
            \textbf{Aug. dataset} & U-Net & 0.959 &  0.120 & 68.130 & 0.967 & 0.950 & \textbf{6.020} & 1.500 \\
            & Segnet & 0.935 & 0.076 & \textbf{69.780} & \textbf{0.975} & \textbf{0.992} & 4.110 & 2.000 \\
            & PSPNet & 0.911 & 0.522 & 60.910 & 0.928 & 0.967 & 5.620 & \textbf{1.200} \\
            & DeepLab v3  & \textbf{0.956} & 0.100 & 64.670 & 0.943 & 0.989 & 5.380 & 4.300  \\
            & Pix2Pix &   0.891 & 0.531 & 59.600 & 0.921 & 0.960 & 4.230 & 6.900  \\
            \bottomrule
            
        \end{tabularx}
        \label{tab:segmentation-metrics}
    \end{center}
\end{table*}

\begin{table*}[htbp]
    \caption{Results comparison for forearm vein \& antecubital region localization \& angle prediction}
    \begin{center}
        \begin{tabularx}{\textwidth}{YYYY}
            \toprule
            \textbf{Model} & \textbf{Multitask loss} & \textbf{MSE} & \textbf{MAE} \\ 
            \midrule
            Modified U-Net \& Dynamic Range Quantization & $\textbf{0.4} \pm \textbf{0.2}$ & $\textbf{42.0} \pm \textbf{6.0}$ & $\textbf{57.7} \pm \textbf{7.3}$ \\
            Base Modified U-Net & $0.4 \pm 0.2$ & $44.6 \pm 6.6$ &  $59.5 \pm 4.7$ \\
            Modified U-Net \& Integer Quantization with Float Fallback  & $0.41 \pm 0.2$ & $58.9 \pm 8.6$ & $73.0 \pm 9.4$ \\
            Modified U-Net \& Full Integer Quantization &  $0.45 \pm 0.2$ & $60.1 \pm 8.3$ & $72.0 \pm 9.7$ \\
            Modified U-Net \& Float 16 Quantization & $0.55 \pm 0.67$ & $76.3 \pm 7.2$ & $82.9 \pm 1.6$ \\
            \bottomrule
        \end{tabularx}
        \label{tab:regression-metrics}
    \end{center}
\end{table*}

For the augmented dataset, several metrics were affected heavily due to the modifications applied to augmented instances. This was an important aspect in the present research since any portable vein finder should also work in more challenging environments than the one where the base dataset was collected. Therefore, we decided to continue the work with the U-Net architecture. 

\subsection{Antecubital Fossa Localization}

\begin{figure}[ht]
  \centering
  \centering
  \subfloat[Graphical User Interface developed with PyQT]{\includegraphics[width=0.50\textwidth]{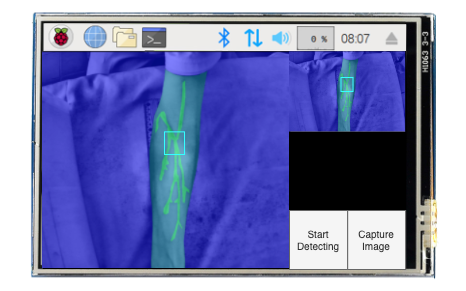}\label{fig:interface}}
  \hfill
  \subfloat[Final printed vein finder]{\includegraphics[width=0.40\textwidth]{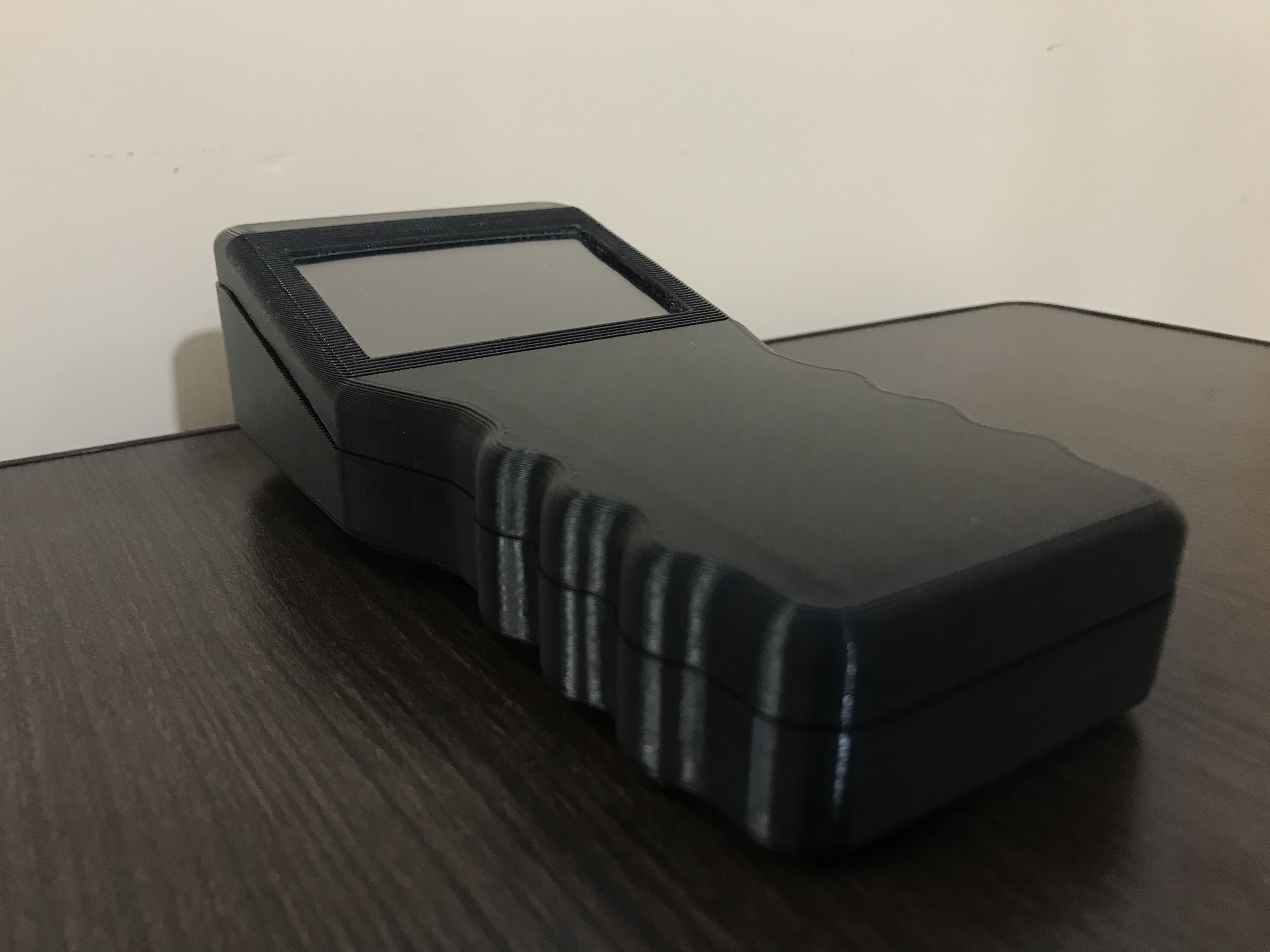}\label{fig:printed-device}}
  \caption{Final prototype printed in 3D}
  \label{fig:Final end device}
\end{figure}

The final model performance was measured considering the metrics MSE and MAE, as defined in Equations \ref{eq:mse} and \ref{eq:mae}, respectively. These results are shown in Table \ref{tab:regression-metrics}, from where we could identify that the compression method Dynamic Range Quantization was the best for the model in terms of precision. Consequently, this was the selected model to be deployed in the end device. The final version of the Graphical User Interface developed with PyQT with the final compressed model inside is shown in Figure \ref{fig:interface}. Finally, the final printed device is shown in Figure \ref{fig:printed-device}. 

\vspace{-10mm}

\section{Conclusions and Future Work}
\label{sec:conclusion}
In this study, we addressed the challenges associated with venipuncture by proposing a comprehensive solution that combines Near Infrared (NIR) imaging and deep learning (DL) techniques for precise vein localization in the antecubital fossa. The significance of accurate vein assessment before intravenous catheterization cannot be understated, especially for patients with low visible veins due to various factors such as fluid retention, age, obesity, dark skin tone, or diabetes. Our proposal comprehends three principal contributions. We introduced a novel dataset comprising 2,016 NIR images of arm veins with limited visibility, accompanied by meticulous annotations that include ground truth images, bounding boxes, centroids, and angle information for precise antecubital fossa identification. 

Furthermore, we devised and compared five different deep learning-based semantic segmentation models, ultimately selecting the most suitable one for antecubital fossa localization and direction prediction. Thirdly, the integration of this model into a compact vein finder device, through rigorous testing of various microcomputers and quantization methods, underlined its feasibility and efficiency in real-world applications. The experimental results demonstrated that the compressed model utilizing Dynamic Range Quantization, deployed on a Raspberry Pi 4B, achieved optimal performance in terms of execution time and precision balance. This achievement, with an execution time of 5.14 frames per second and an Intersection over Union (IoU) of 0.957, showcased the potential of our approach in a resource-constrained and cost-effective portable device.

For future work, other imaging modalities should be combined. Moreover, we highlight the importance of recognizing the median cubital vein, as well as other vascular structures shown in Figure \ref{fig:vein-distribution}, in future computer vision-based vein detectors. In addition, suitable vein recommendations according to a given intravenous procedure should be also considered in future research to enhance venipuncture procedures and patient care.

\bibliographystyle{splncs04}
\bibliography{bibliography}

\end{document}